\documentclass[preprint,review,12pt]{elsarticle}

\usepackage{graphicx}
\usepackage{amssymb}
\usepackage{lineno}

\biboptions{comma,square, numbers, sort&compress}
\journal{Physica A}

\begin{document}

\begin{frontmatter}
  \title{Synchronization time in a  hyperbolic dynamical system with long-range interactions}

  \author{Rodrigo F. Pereira\fnref{uepg}}
  \ead{pereira.rf@gmail.com}
  
  \author{Sandro E. de S. Pinto\corref{cor1}\fnref{uepg}}
  \ead{desouzapinto@pq.cnpq.br}
  \cortext[cor1]{Corresponding author}

  \author[ufpr]{Sergio R. Lopes}

  \address[uepg]{Departamento de F\'isica,Universidade Estadual de Ponta Grossa, 84030-900, Ponta Grossa, PR, Brazil}
  \address[ufpr]{Departamento de F\'isica, Universidade Federal do Paran\'a, 81531-990, Curitiba, PR, Brazil}

  \begin{abstract}
    We show that the threshold of complete synchronization in a lattice of coupled non-smooth chaotic maps is determined by linear stability along the directions transversal to the synchronization subspace. We examine carefully the sychronization time and show that a inadequate observation of the system evolution leads to wrong results. We present both careful numerical experiments and a rigorous mathematical explanation confirming this fact, allowing for a generalization involving hyperbolic coupled map lattices.
  \end{abstract}


\end{frontmatter}

\linenumbers

The possibility of synchronizing chaotic dynamics has been harnessed in a large number of systems of physical interest \cite{BKpikovsky.2001, PR.v366.p1.y2002}, like coupled Josephson junctions \cite{PRL.v76.p404.y1996} and lasers \cite{PRL.v72.p2009.y1994}. Although there have been identified different types of chaos synchronization, we shall concentrate on the so-called amplitude or complete synchronization, for which all dynamical variables undergo the same time evolution \cite{PRL.v64.p821.y1990}. The essential dynamics involved in the process of chaos synchronization lies on the low-dimensionality of the subspace (in the phase space of the system) in which synchronized motion sets in. 

For example, if we consider a lattice of $N$ coupled oscillators, each of them represented by a vector field of $D$ dimensions, where typically $D \ll N$, the synchronized state belongs to a $D$-dimensional subspace of the $ND$-dimensional phase space. In order for this synchronized state to exist the coupling among oscillators takes on a suitable form \cite{Chaos.v7.p520.y1997}. Whether or not this synchronized state is stable, however, is a more difficult question, since it involves the analysis of infinitesimal displacements from the synchronized state along all $(N-1)D$ directions transversal to the synchronization subspace \cite{PRE.v58.p347.y1998}. The stability condition of the synchronized orbit with respect to transversal perturbations can be obtained from the negativeness of the largest transversal Lyapunov exponent.

In this paper we consider a coupled chaotic map lattice (CML) in which the coupling prescription is linear and non-local, for it takes into account the distance between maps along the lattice. Such non-local couplings appear in many problems of physical \cite{PRE.v75.p036209.y2007} and biological interest \cite{PRL.v74.p3297.y1995}. We suppose that the coupling strength decreases with the lattice distance as a power-law, which characteristic exponent can take on any non-negative value \cite{PRE.v65.p056209.y2002}. The loss of transversal stability of the synchronized state, as the coupling parameters are varied, was found in such power-law couplings, with help of the largest transversal Lyapunov exponent, for a number of chaotic maps \cite{PhysicaD.v206.p94.y2005, PRE.v68.p067204.y2003}. In the particular case of maps with constant eigenvalues of the Jacobian matrix (piecewise-linear chaotic maps) we obtained analytical results for the loss of transversal stability of the synchronized state which agree with the numerical simulations \cite{PRE.v68.p45202.y2003}. Such CML's represent hyperbolic dynamical systems (see text below), what enables us to use powerful mathematical tools like ergodicity and global shadowing of numerically generated orbits \cite{BKkatok.1996}.

On the other hand, in a recent paper there was argued that in the special case of coupled non-smooth discontinuous maps the synchronization transition would not be given by the largest transversal exponent, but rather by a different approach taking into account finite distances from the synchronized state \cite{PhysicaD.v208.p191.y2005}. To investigate this apparent contradiction we considered in this paper the transient behavior of the non-synchronized orbits for coupled piecewise linear maps. Our results show that the analytical results of Ref. \cite{PRE.v68.p45202.y2003} (using linear transversal stability of the synchronized state) {\it hold for both smooth and non-smooth maps}, the numerical results being strongly affected by many factors as the large transient time and the choice of initial conditions. Due to these factors, the time it takes to achieve convergence to the synchronized state may be extremely large, what may lead to wrong conclusions about the stationary state of the system. Motivated by this problem, we investigated the validity of the transversal linear stability analysis in a class of hyperbolic CML's, using periodic-orbit theory to unveil the role of the unstable orbits embedded in the synchronized state \cite{PRA.v37.p1711.y1988,PRE.v56.p4031.y1997,Chaos.v17.p023131.y2007}. 

The CML we consider in this work can be written in the explicit form of a $N$-dimensional dynamical system 
\begin{equation}
\textbf{x}_{n+1}= (\textbf{1}+\textbf{C})\textbf{F}(\textbf{x}_n)\equiv \textbf{B}\textbf{F}(\textbf{x}_n),
\label{coupledmap}
\end{equation}
\noindent where the components of $\textbf{x}_n={(x_n^{(1)},x_n^{(2)},\ldots,x_n^{(N)})}^T$ denote the state variable attached to the map located at the site $i=1,\ldots,N$ at time $n=0,1,\ldots$. If the uncoupled maps are written as $x \mapsto f(x)$ we can write $\textbf{F(\textbf{x})}={[f(x^{(1)}),f(x^{(2)}),\ldots,f(x^{(N)})]} ^T$. Moreover, the coupling prescription is represented by the matrix $\textbf{C}$, and $\textbf{1}$ is the identity matrix. 

In the following we consider the generalized Bernoulli map $f(x)=\beta x$, $(mod\;1)$, where $x \in [0,1)$ and $\beta > 1$, such that the isolated map generates a strongly chaotic orbit. When these piecewise-linear maps are coupled according to Eq. (\ref{coupledmap}), in order to ensure that $x_n^{(i)} \in [0,1)$, the elements of $\textbf{B}$ must satisfy the following necessary and sufficient conditions: $B_{ij}\geq 0$, and $0 \leq \sum_{j=1}^N B_{ij} \leq 1$, for all $i,j = 1, 2, \ldots N$ \cite{PRE.v72.p037206.y2005}. Moreover, we use a symmetric coupling matrix with elements 
\begin{equation}
C_{ij}=\varepsilon \eta^{-1} \left\lbrack r_{ij}^{-\alpha}(1-\delta_{ij}) -\eta\delta_{ij} \right\rbrack,
\end{equation}
\noindent where $r_{ij}=\mbox{min}_{l\in\mathbb{Z}}|i-j+lN|$ is the minimum lattice distance between the sites $i$ and $j$ (with periodic boundary conditions), $\eta=2\sum_1^{N'} r^{-\alpha}$, with $N'=(N-1)/2$ (which requires $N$ odd), and the coupling strength satisfies $0 \le \varepsilon \le 1$ due to the constraints on $B_{ij}$. The effective range of interactions is represented by $\alpha\ge 0$ such that the limits $\alpha=0$ and $\alpha\rightarrow \infty$ correspond, respectively, to global (mean field) and local (first neighbors) coupling prescriptions. 

A completely synchronized state is the chaotic orbit for which $x_n^{(1)}=\ldots=x_n^{(N)}$, and which is a solution of Eq. (\ref{coupledmap}). Since the Jacobian ${\bf DF} = \beta{\bf B}$ is a circulant matrix, its eigenvalues can be analytically obtained as $$\Lambda^{(k)} = \beta [(1-\varepsilon) + (\varepsilon/\eta) b^{(k-1)}],$$ where 
\begin{equation}
\label{eigenvalues}
b^{(k)} = \sum_{m=1}^{N'} \frac{1}{m^\alpha} \cos\left(\frac{2\pi k m}{N}\right), \quad (0\leq k < N)
\end{equation}
\noindent such that the Lyapunov spectrum ${\{\lambda_i\}}_{i=1}^N$ can be derived \cite{PRE.v68.p45202.y2003}. The stability threshold of the synchronized state with respect to infinitesimal transversal displacements, obtained by imposing $\lambda_2=0$, gives two curves in the parameter plane ($\varepsilon$ {\it versus} $\alpha$): (i) $\varepsilon_c'(\alpha,N) = \mbox{min}\{\varepsilon_{up}(\alpha,N),1\}$; and (ii) $\varepsilon_c(\alpha,N) = \mbox{min}\{\varepsilon_{lo}(\alpha,N),1\}$, where we defined
\begin{eqnarray}
\varepsilon_{up}(\alpha,N) & = & (1 + {\beta}^{-1}) {[1 - (b^{(N')}/\eta)]}^{-1}, \\
\varepsilon_{lo}(\alpha,N) & = & (1 - {\beta}^{-1}) {[1 - (b^{(1)}/\eta)]}^{-1}.
\end{eqnarray}

In order to check the validity of these analytical conditions for the threshold of transversal stability we have made careful numerical experiments using the same criteria as proposed in Ref. \cite{PhysicaD.v208.p191.y2005} (where it has been claimed that those conditions would hold only for coupled continuous maps). Accordingly, we choose initial conditions $x_0^{(i)}$ uniformly distributed in the interval $[0,1)$ \footnote{We initialized the random number generator {\tt ran1} from Ref. \cite{BKpress.1992} with always the same the seed ($-28937104$)}. The CML is firstly iterated for a transient time of $T_w = 10^w \times N$ times and further iterated by more $T = 10^3 \times N$ times. As a numerical diagnostic of complete synchronization we computed the following quantity 
\begin{equation}
\label{order}
R = \sum_{n,i} \frac{1}{NT} \left\vert x_n^{(i)}-\left( \frac{1}{N} \sum_{j} x_n^{(j)} \right) \right\vert,
\end{equation}
\noindent which is essentially a mean deviation from the lattice-averaged amplitude. The resulting dynamical state is considered as being completely synchronized if $R < 10^{-8}$. 
\begin{figure}[htb]
  \includegraphics[width=0.75\columnwidth,clip]{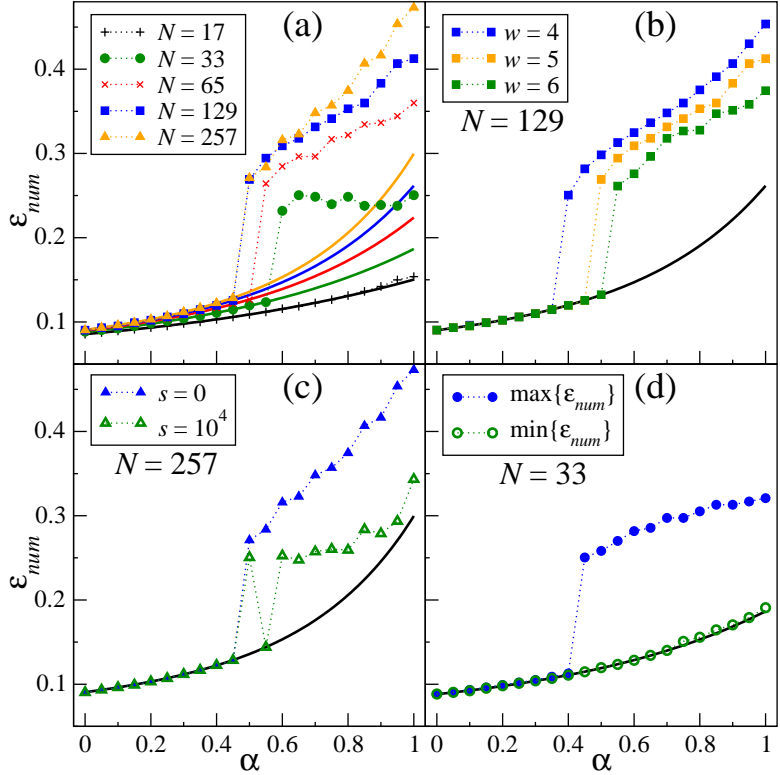}
  \caption{\label{FIGcurves} (color online) Values of the coupling strength at the onset of transversal stability loss of the synchronized state, as a function of the effective coupling range. We used $\beta = 1.1$ and (a) different lattice sizes; (b) different transient times $T_w$, for a fixed lattice size; (c) different distributions of initial conditions, for $N = 257$; (d) different initial conditions, for $N = 33$ and a fixed transient time. The solid lines represent the analytical results from linear transversal stability of the synchronized state.}
\end{figure}
In the coupling parameter space we keep $\alpha$ constant and sweep through the values of $\varepsilon \in [0,1]$. The value corresponding to the synchronization threshold, denoted as $\varepsilon_{num}$, is obtained from bisection as $\varepsilon_{num}=(\varepsilon_s+\varepsilon_d)/2$, where $\varepsilon_s$ and $\varepsilon_d$ are, respectively, the last value corresponding to a synchronized state and the first value for a non-synchronized one. The numerical value of $\varepsilon_{num}$ is turned more accurate from refining the increment mesh and repeating the process, until $(\varepsilon_s-\varepsilon_d) \leq 10^{-3}$. 

The results of this numerical procedure, for the case $\beta=1.1$, are depicted in Figure \ref{FIGcurves}, where we show the value of the coupling strength at the synchronization threshold as a function of $\alpha$. In Fig. \ref{FIGcurves}(a) we show how the numerically determined critical value increases with $\alpha$ for different lattice sizes $N$, the transient time being different for each choice, using $w = 5$. The solid lines correspond to the analytical condition derived in Ref.\cite{PRE.v68.p45202.y2003} (and that depend on the lattice size as well). In fact, as the lattice size $N$ increases, the numerical values of $\varepsilon_{num}$ may no longer match the analytically predicted values, if $\alpha$ is large enough \cite{PhysicaD.v208.p191.y2005}. This does not mean, however, that the analytical value of $\varepsilon_{c}$ is not valid in those cases, but rather that {\it the numerical simulations have not been performed using a transient long enough}. To show the influence of the transient time in the results, we show in Fig. \ref{FIGcurves}(b) the dependence of $\varepsilon_{num}$ with $\alpha$ for a fixed lattice of $N = 129$ sites by changing the parameter $w$. By increasing the transient time the numerically obtained values for the synchronization threshold agree better with those derived from transversal linear stability. The same conclusions were obtained using other lattice sizes as well. These results suggest that the analytical result for $\varepsilon_{c}$ remains valid, as long as we use sufficiently long transient times, in contrast with Ref. \cite{PhysicaD.v208.p191.y2005}.

Another factor that affects the accuracy of numerical results for the threshold of synchronization is that a distribution of initial conditions over the interval $[0,1)$ should respect the natural measure of the chaotic orbit, because we are assuming the coupling between typical oscillators, which are characterized by trajectories in the steady-state system, {\it i. e.}, trajectories that satisfy the invariant density of the system. While for integer values of $\beta$ the natural measure is uniform, this is no longer valid for fractional $\beta$, and small errors may be introduced if we choose initial conditions with a uniform probability distribution. In order to overcome this problem we iterated each map $s$ times before starting coupling them according to Eq. (\ref{coupledmap}) (this transient time should not be confused with the transient time $T_w$ we compute after having started coupling the maps). In Figure \ref{FIGcurves}(c) we compare the results of two simulations: for the line with filled triangles we used initial conditions uniformly distributed along $[0,1)$, without discarding any transients $(s = 0)$; whereas the line with open triangles was obtained from initial conditions chosen with respect to a numerical approximation of the natural measure, the latter having being obtained from a transient time of $s = 10^4$ iterations. The results obeying the natural measure of the uncoupled oscillators are more likely to agree with the analytical results since, after the synchronized state sets in, the corresponding orbit must follow this natural measure. As will be formally discussed below, in the limit $n\rightarrow\infty$, the results are independet of the (typical) initial distribution of trajectories. But for finite time intervals, time synchronization can depend on such distribution. This dependence is due to local dynamics and form of coupling.

In order to analyze the dependence of $\varepsilon_{num}$ on the initial conditions, we performed extensive numerical simulations with an ensemble of $5000$ identically prepared CML's, each of them with a different initial condition [Fig. \ref{FIGcurves}(d)] and the same transient time ($w = 5$). We observed different values of $\varepsilon_{num}$ for each initial condition, provided $\alpha$ is large enough. Instead of showing each of them (what would turn the figure too much loaded with symbols) we represented in Fig. \ref{FIGcurves}(d) only those numerical values of $\varepsilon_{num}$ that are closest (open circles) and farthest (filled circles) with respect to the analytical value (full line). Note that synchronization of one typical trajectory implies global stability of the synchronized state, since the system is hyperbolic. A chaotic invariant set $\Omega$ is hyperbolic if the following conditions are fulfilled: (i) the tangent space at each point ${\bf x} \in \Omega$ can be decomposed in two invariant subspaces (a stable and an unstable one) with constant dimensions; (ii) these subspaces always intersect transversely (i.e., they cannot present tangencies); and (iii) this decomposition is consistent under the dynamics in $\Omega$ generated by ${\bf F}$ \cite{BKkatok.1996}. For coupled generalized Bernoulli maps the set $\Omega$ is the $N$-torus ${[0,1)}^N$ and the Jacobian matrix ${\bf DF}$ has constant entries and does not depend on ${\bf x} \in \Omega$, thus the dimension of the invariant subspaces is constant everywhere [condition (i)]. Thanks to this particular form of the Jacobian its eigenvectors (which span the invariant subspaces) are everywhere orthogonal [condition (ii)]. Let ${\bf u}$ be any of such eigenvectors: under the dynamics of ${\bf F}$ it follows that ${\bf u}$ is mapped to a vector along the same direction [condition (iii)]. Hence the set $\Omega$ is a hyperbolic structure for ${\bf F}$.

It is possible to understand, from a general point of view, the causes of the strong dependence of the synchronization threshold results on the transient time and the initial conditions. These causes are not restricted to coupled piecewise-linear maps as ours, but are rather generic for hyperbolic CML's. We can extend our conclusions to a CML given by Eq. (\ref{coupledmap}) where the coupling prescription keeps invariant the phase space $\Omega = {[0,1)}^N$, and for which $$\mathcal{S}=\{\mathbf{x}\in\Omega: x^{(1)}=\cdots=x^{(N)}\}$$
is the one-dimensional invariant synchronization manifold defined by the corresponding state. We consider a $\Delta$-neighborhood of $\mathcal{S}$ as the set of points whose distances from the $\mathcal{S}$ do not exceed $\Delta$: $\Sigma_\Delta = \{\mathbf{x}:{\sf d}(\mathbf{x},\mathcal{S})\leq \Delta\}$, where ${\sf d}$ is a suitably defined distance on the metric space $\Omega$. We define $\Sigma \equiv \lim_{\Delta\rightarrow 0} \Sigma_\Delta$ as a linear neighborhood of $\mathcal{S}$. Accordingly $\Gamma = \Omega - \Sigma$ is the phase space region, except the linear neighborhood of the synchronization manifold.

We can speak of the global dynamics generated by the coupled map lattice $\textbf{x}_{n+1}= \textbf{B}\textbf{F}(\textbf{x}_n)$ in terms of their periodic points. In this spirit we denote $\mathbf{x}_j(p)$ the $j$th fixed point of the $p$-times iterated vector function $\textbf{B}\textbf{F}(\textbf{x}_n)$. The $i$th eigenvalue of the Jacobian matrix of ${\textbf{B}\textbf{F}}^{[p]}(\textbf{x}_n)$, evaluated at this point, is written as $\Lambda_i(\mathbf{x}_j(p))$, such that $|\Lambda_1(\mathbf{x}_j(p))|\geq \cdots \geq |\Lambda_N(\mathbf{x}_j(p))|$.

Let us consider a subset of the phase space, $A\subset \Omega$, with natural measure $\mu(A)$. Note that, by construction, we have  $\mu(\Omega)=1$. For hyperbolic  systems satisfying the Axiom-A \footnote{A hyperbolic system satisfying Axiom-A must be also mixing. This condition is fulfilled if the system possesses a dense set of unstable periodic orbits embedded in the phase space \cite{BKkatok.1996}.} the natural measure of such subset can be obtained from the unstable periodic points embedded in it as \cite{PRA.v37.p1711.y1988} 
\begin{equation}\label{EQmeasureupos}
\mu(A) = \lim_{p\rightarrow\infty} \sum 1/L_j(p),
\end{equation}
\noindent where $L_j(p) = \prod_{i=1}^{d_u}|\Lambda_i(\mathbf{x}_j(p))|$ ($d_u$ is the largest integer such that $|\Lambda_{d_u}(\mathbf{x}_j(p))|>1$) and the sum sweeps over all $\mathbf{x}_j(p)\in A$. The exploitation of this identity is the object of periodic-orbit theory, that has been used for a number of theoretical investigations on the properties of chaotic dynamical systems \cite{PRE.v56.p4031.y1997,Chaos.v17.p023131.y2007}. For generalized Bernoulli maps $\beta x$ (mod $1$) and a linear coupling, the Jacobian matrix has constant entries and thus do not depend on the orbit points, \textit{i.e.}, all the unstable periodic orbits have the same eigenvalue spectra (consequently $L_j(p) = L(p)$ for all $j$), and the natural measure is $\mu(A) = \lim_{p\rightarrow\infty} N_A(p)/L(p)$, where $N_A(p)$ is the number of period-$p$ points contained in the subset $A$ of $\Omega$.

A byproduct of the periodic-orbit theory is that the (linear) transversal stability of the synchronization manifold can be studied either from the natural measure of a typical chaotic orbit (by the second largest Lyapunov exponent) or from the atypical measure generated by the unstable periodic orbits. In particular, with respect to the period-$p$ orbit the threshold of transversal stability of the synchronization manifold can be obtained from the condition $|\Lambda_2(\mathbf{x}_j(p))| = 1$ for all $\mathbf{x}_j(p) \in {\cal S}$. As the period $p$ goes to infinity we expect an increasingly better agreement of this result with that obtained by using the second largest transversal Lyapunov exponent (or $\lambda_2 = 0$). For a given $\alpha$ and values of the coupling strengths such that $\varepsilon_{lo}(\alpha) < \varepsilon < \varepsilon_{up}(\alpha)$, the natural measure of the subset $A$ is 
\begin{equation}
\mu(A) = \lim_{p\rightarrow\infty} N_A(p)/\beta^p.
\end{equation}

Taking $A$ to be the linear neighborhood of the synchronization manifold, $\Sigma$, there follows that the number of orbits in this neighborhood is $N_{\Sigma} = \beta^p - 1$ for integer $\beta$ (if $\beta$ is fractional, as in the numerical simulations of the previous section, $N_p \rightarrow \beta^p$ for $p\gg 1$) and the corresponding natural measure is given by 
\begin{equation}
\mu(\Sigma) = \lim_{p\rightarrow\infty} (\beta^p - 1)\beta^{-p} = 1,
\end{equation}
\noindent demonstrating that the linear neighborhood of the synchronization manifold $\mathcal{S}$ is the asymptotic state of any typical initial condition (in the sense that the set of initial conditions that do not converge to $\Sigma$ has zero Lebesgue measure). This result is obtained for the parameter regime in which the synchronized state is locally stable and, therefore, any trajectory in $\Sigma$ converges exponentially to $\mathcal{S}$ at a rate $\lambda_2<0$. An immediate consequence of this result is that the natural measure outside the linear neighborhood is zero since, using the fact that the natural measure is ergodic, we have $\mu(\Gamma) = \mu(\Omega) - \mu(\Sigma) = 0$. 

Given that almost all initial conditions outside the synchronization manifold eventually asymptote to it, we may well ask why sometimes it takes so long for this convergence to be observed in numerical experiments. As we saw previously, this long transient time may even be mistaken as a effect of non-convergence. The answer lies in the properties of the horseshoe-like invariant chaotic set embedded in $\Gamma$. This set is non-attracting since almost all initial conditions in $\Gamma$ converge to ${\cal S}$ as the time goes to infinity. 

Let $\tilde{\rho}_n(\mathbf{x})$ be the density of trajectories arounf $\mathbf{x}$ at time $n$, so that
\begin{equation}
  \mu(A) = \lim_{n\rightarrow\infty}\int_A \tilde{\rho}_n(\mathbf{x})\mbox{d}\mathbf{x}
\end{equation}
for any typical $\tilde{\rho}_0(\mathbf{x})$. The above result is independent of the specific form of $\tilde{\rho}_0$ in the limit $n\rightarrow\infty$. The possibility of expanding $\tilde{\rho}_0$ in terms of the eigenfunctions of the Perron-Frobenius operator justifies such independence of $\mu(A)$, since the invariant density $\rho(\mathbf{x})$ of the system is associated with the largest eigenvalue (which is not degenerate) of that operator \cite{BKbeck.1995}. Note that for finite time intervals, the convergence of the invariant density depends on the coefficients of expansion of $\tilde{\rho}_0(\mathbf{x})$ on the basis of eigenfunctions of the Perron-Frobenius operator. The results in Figure \ref{FIGcurves}(c) are evidence of the assertion of the previous sentence.

\begin{figure}[htb]
  \includegraphics[width=0.75\columnwidth,clip]{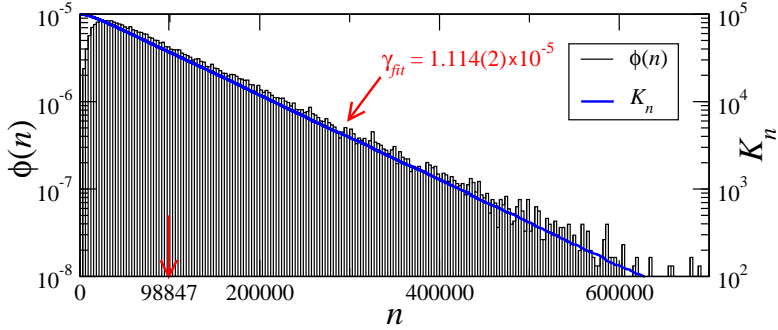}
  \caption{\label{FIGtimes} (color online) Line: decay in the number of trajectories in the unsynchronized state for $N=17$, $\beta=1.10$, $\alpha=0.80$ and  $\varepsilon=0.17$ (slighty above the critical curve in Figure \ref{FIGcurves}). Histogram: synchronization times distribution for same lattice parameters. The exponential tail of such distribution is given by the escape rate of the saddle embedded in $\Gamma$. The average synchronization time, $\langle n \rangle$ = 98847, is aproximately given by $\gamma_{fit} ^{-1}$.}
\end{figure}

However, the measure generated by chaotic orbits whose initial conditions are uniformly distributed over of an open region $B$ of the phase space $\Omega$ decays exponentially with time with escape rate $\gamma$,
\begin{equation}\label{EQmeasuredecay}
  \int_B \tilde{\rho}_{n+m}(\mathbf{x})\mbox{d}\mathbf{x} = e^{-\gamma n}\int_B \tilde{\rho}_m(\mathbf{x})\mbox{d}\mathbf{x}
\end{equation}
with $m\gg 1$. For hyperbolic systems it can be shown that the escape rate is also obtained in terms of unstable periodic points of the saddle according to
\begin{equation}
  \lim_{p\rightarrow\infty}e^{-\gamma p}\sum_{\mathbf{x}_j(p)\in B} 1/L_j(p) = 1,
\end{equation}
\noindent where in the sum we consider only the periodic orbits of the horseshoe-like set $B$ outside the synchronization manifold \cite{PRA.v37.p1711.y1988}. Hence, if one picks up at random an initial condition off the synchronization manifold, the distribution of the transient times is likely to be exponential, with a characteristic  exponent dependent on the escape rate $\gamma$. This is illustrated in Figure \ref{FIGtimes}, in which the distribution of synchronization times (transient time intervals), $\phi(n)$, is indicated for a typical realization of the network, with $N=17$. Figure \ref{FIGtimes} also points out a numerical estimate of the density measure decay of the unsynchronized state. To obtain this estimate, we cover $\Gamma$ with $K_0$ uniformly distributed initial conditions and, at each time instant $n$, we count the number $K_n$ of trajectories that remain in $\Gamma$ under the system evolution. The temporal decay of $K_n$ provides the escape rate of $\Gamma$ since $K_n/K_0 \approx \int_\Gamma\tilde{\rho}_n(\mathbf{x})\mbox{d}\mathbf{x}$, for $K_0\gg 1$. Note that the exponential decay of the curves is the same in Figure \ref{FIGtimes}. The mean time of synchronization (or the mean life of the chaotic saddle), given by
\begin{equation}\label{EQmeantime}
  \left<n\right> = \int_0^\infty n\phi(n) \mbox{d}n \approx \frac{1}{\gamma},
\end{equation}
is indicated by a red arrow on the $x$-axis. The rightmost term in Eq. (\ref{EQmeantime}) is obtained by supposing $\phi(n)\propto e^{-\gamma n}$ -- which is typical for chaotic saddles and is verified in Fig. \ref{FIGtimes}.

\begin{figure}[htb]
  \includegraphics[width=0.75\columnwidth,clip]{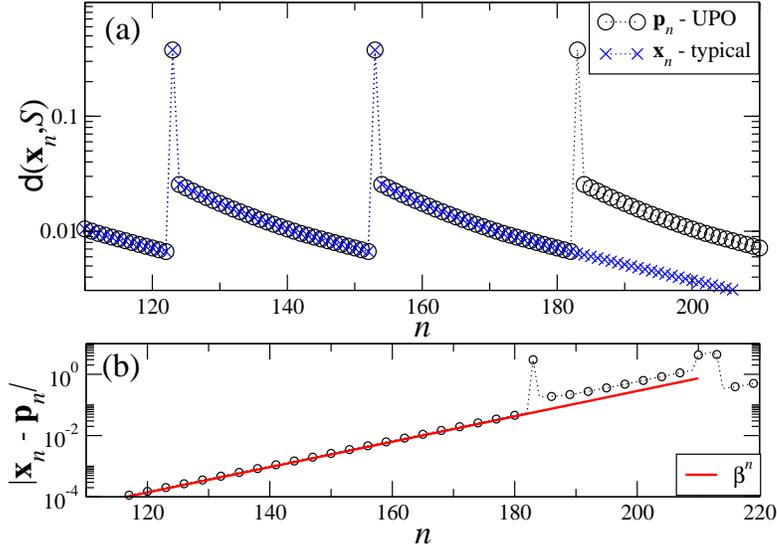}
  \caption{\label{FIGshadow} (color online) (a) Series of a typical trajectory, $\mathbf{x}_n$, in the vicinity of a periodic orbit $\mathbf{p}_n\in\Gamma$. The trajectory follows the periodic orbit for a few periods, and then escapes from the saddle and reach the synchronized state. (b)The distance between the trajectory and the periodic orbit as a function of time $n$. Trajectory moves away from the UPO at a rate given by the unstable eigenvalue of the orbit.}
\end{figure}

If the initial condition is too close to an unstable periodic orbit (or its stable manifold) of $B$ it would stick to it for some time-span and hence it takes a very long time for such a trajectory to approach the synchronization manifold, as illustrated in Fig. \ref{FIGshadow}. This seems to occur very often if we use fractional values of $\beta$, like in the numerical simulations we shown in this paper. 

We show in Figure \ref{FIGshadow} a situation in which the trajectory escapes from the neighborhood of an orbit with small period ($\sim 30$), and instantly access the synchronized state. However, since there is a chaotic saddle in $\Gamma$, there is a dense set of unstable periodic points in the unsynchronized state, and a trajectory can wander between different UPO's before escaping to the synchronized state.

\begin{figure}[htb]
  \includegraphics[width=\hsize, clip]{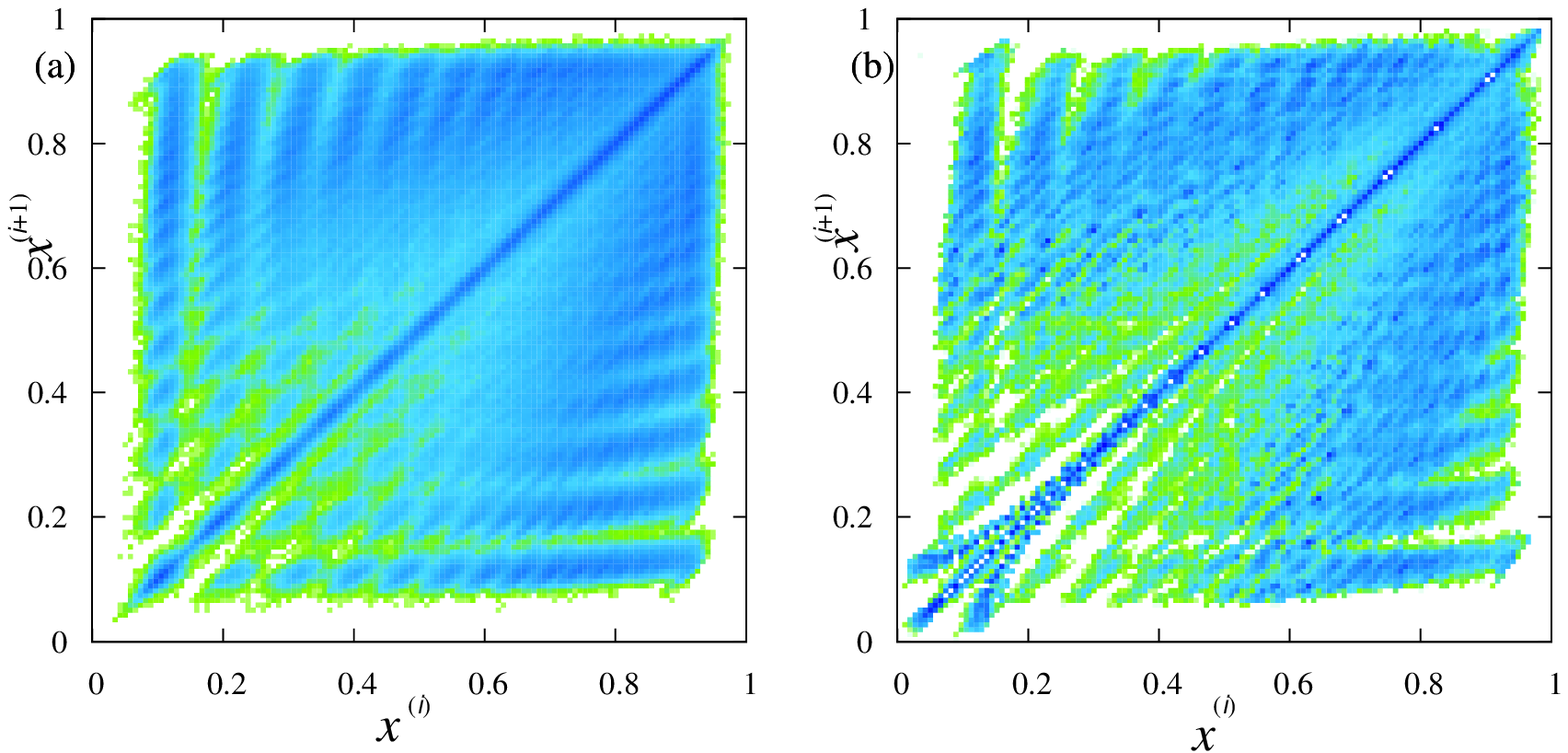}
  \caption{\label{FIGsaddle} (color online) (a) Density projection of the chaotic saddle that is contained in $\Gamma$, for typical trajectories. The color scale indicates the density: the blue color indicates regions most visited and the green color indicates the least visited. (b) Same as previous, but using all unstable orbits with the period less than or equal to 35. The simulations were performed with the following parameters: $N=17$, $\beta=1.10$, $\alpha=0.80$ and $\varepsilon=0.17$.}
\end{figure}

To demonstrate that the behavior shown in Figure \ref{FIGtimes} is typical, we present in Figure \ref{FIGsaddle} two numerical estimates for the invariant density of the chaotic saddle in the unsynchronized state. In Figure \ref{FIGsaddle}(a), which is obtained from typical trajectories, the invariant density is shown in a projection on the $x^{(i)}\times x^{(i+1)}$ plane (due to symmetry the network, the value of $i$ is irrelevant). The most visited regions in that figure are represented by symbols whose color is dark blue and the least visited by  light green symbols. The estimate for the density in terms of UPO's is shown in Figure \ref{FIGsaddle}(b). For this case, all the unstable  orbits with the period less than or equal to 35 are considered. Comparing the two figures is apparent that the structures in Figure \ref{FIGsaddle}(a) are supported by the orbits in Figure \ref{FIGsaddle}(b). The processing time was the limiting factor for choosing the value of the period in the simulations shown in Fig. \ref{FIGsaddle}(b)\footnote{For $p = 28$ there are around 10000 periodic points in $\Gamma$. However, for $p = 35$ there are around 10000000 points.}. The Table \ref{TABdiff} shows the mean differences between the values of projections of $\rho(\mathbf{x})$ and $\rho_p(\mathbf{x})$, the density values obtained by calculation of all points in $\Gamma$ with period $q\leq p$. This calculation was done in a grid of $128\times 128$ boxes, each one as the same size. We also found that the average difference decreases as the period increases, this is according to reference \cite{PRL.v79.p649.y1997}. When the period tends to infinity, we obtain the exact measure of the saddle \cite{PRA.v37.p1711.y1988}.

\begin{table}[htb]
  \centering
  \begin{tabular}{|c|c|}
    \hline Period $p$ & $\langle \rho(\mathbf{x}) - \rho_p(\mathbf{x})\rangle\times 10^4$ \\
    \hline 29 & 1.164\\
    \hline 30 & 1.157\\
    \hline 31 & 1.153\\
    \hline 32 & 0.690\\
    \hline 33 & 0.499\\
    \hline 34 & 0.458\\
    \hline 35 & 0.425\\ \hline 
  \end{tabular}
  \caption{\label{TABdiff} The mean difference $\langle \rho(\mathbf{x}) - \rho_p(\mathbf{x})\rangle$, magnified by a $10^4$ factor, in a grid of $128\times 128$ boxes, each one as the same size.}
\end{table}

From the view of the structure of the chaotic saddle provided by Figure \ref{FIGsaddle}, it is easy to understand how the distribution of initial conditions affects the results for simulations with finite time. For the case we are considering, the Bernoulli map with $\beta=1.10$, the density of the natural measure in the unit interval is mainly concentrated near zero. In Figure \ref{FIGsaddle}(b) the chaotic saddle is less dense at the bottom left than in the upper right corner, {\it i.e.}, the initial conditions close to zero spend less time to synchronize. However, it should be noted that this dependence with the initial distribution of trajectories is a consequence of the finite time simulations because, as shown, such as effects are transient due to the existence of the chaotic saddle in $\Gamma$. Figure \ref{FIGsaddle} shows a high value for the invariant density for specific points along the diagonal that contains the projection of the synchronized state. These points, which are in the chaotic saddle, are those associated with the discontinuity of the local map, $\beta^{-1}$, and the respective pre-images. Thus, the probability of a trajectory in the neighborhood of such points still belongs to the chaotic saddle is very high. Consequently, the behavior illustrated in Figure \ref{FIGshadow}(a) is more likely to occur when a trajectory close to $\mathcal{S}$ is near of a discontinuity (or its pre-images) of the local dynamics. This is the topological explanation for the non-local instabilities in synchronized state, which supports the heuristic argument presented in Reference \cite{PhysicaD.v208.p191.y2005}. Direct analysis of our results, especially those presented in Figures \ref{FIGshadow} and \ref{FIGsaddle}, shows that the divergence between a typical trajectory and locally stable synchronized state is due strictly to the existence of UPO's in the unsynchronized state.

In conclusion, the analytical conditions for the threshold of transversal stability of the synchronized state of coupled piecewise-linear maps are confirmed by numerical experiments as long as we observe the following precautions: (i) the transient time should be chosen as large as possible, (ii) the choice of initial conditions should be done using a probability distribution which best matches the natural measure of the uncoupled oscillators.

Although these computational problems are less likely to occur in coupled smooth maps, they do not invalidate the analytical approach to the transversal stability of coupled non-smooth maps, like piecewise-linear ones. Our analysis indicates that results based on linear analysis of stability of the synchronized state may be valid for both smooth and non-smooth local dynamics. Thus, results as those found in reference \cite{PRE.v76.p027202.y2007} are directly extended to piecewise-linear coupled maps. We have used general arguments valid for hyperbolic CML's so as to prove that the local transversal stability of the synchronized state actually implies the synchronization of {\it all} typical orbits. Finally, since there is the conjecture \cite{PRA.v37.p1711.y1988} that the expression (\ref{EQmeasureupos}) is valid for non-hyperbolic systems, we conjecture that our results are valid for networks whose phase space has no structure hyperbolic.

\bibliographystyle{elsarticle-num}
\bibliography{biblio}

\end{document}